\documentclass[aps,superscriptaddress,prc,twocolumn,nofootinbib]{revtex4}

\usepackage{amsmath,bm}
\usepackage{graphicx}
\usepackage{epstopdf}
\usepackage{subfigure}
\usepackage{epsfig}
\usepackage{amsmath,amssymb,amsfonts}
\usepackage{color}
\usepackage[utf8]{inputenc}
\usepackage{verbatim}
\usepackage{array}

\begin{document}

\newcommand{\vk}{{\vec k}}
\newcommand{\vK}{{\vec K}}
\newcommand{\vb}{{\vec b}}
\newcommand{{\vp}}{{\vec p}}
\newcommand{{\vq}}{{\vec q}}
\newcommand{\vQ}{{\vec Q}}
\newcommand{\vx}{{\vec x}}
\newcommand{\beq}{\begin{equation}}
\newcommand{\eeq}{\end{equation}}
\newcommand{\half}{{\textstyle \frac{1}{2}}}
\newcommand{\gton}{\stackrel{>}{\sim}}
\newcommand{\lton}{\mathrel{\lower.9ex \hbox{$\stackrel{\displaystyle<}{\sim}$}}}
\newcommand{\ee}{\end{equation}}
\newcommand{\ben}{\begin{enumerate}}
\newcommand{\een}{\end{enumerate}}
\newcommand{\bit}{\begin{itemize}}
\newcommand{\eit}{\end{itemize}}
\newcommand{\bc}{\begin{center}}
\newcommand{\ec}{\end{center}}
\newcommand{\bea}{\begin{eqnarray}}
\newcommand{\eea}{\end{eqnarray}}

\newcommand{\beqar}{\begin{eqnarray}}
\newcommand{\eeqar}[1]{\label{#1} \end{eqnarray}}
\newcommand{\pleft}{\stackrel{\leftarrow}{\partial}}
\newcommand{\pright}{\stackrel{\rightarrow}{\partial}}

\newcommand{\eq}[1]{Eq.~(\ref{#1})}
\newcommand{\fig}[1]{Fig.~\ref{#1}}
\newcommand{\eff}{ef\!f}
\newcommand{\alphas}{\alpha_s}

\renewcommand{\topfraction}{0.85}
\renewcommand{\textfraction}{0.1}
\renewcommand{\floatpagefraction}{0.75}

\title{Medium modifications of girth distributions for inclusive jets and $Z^0+{\rm jet}$ in relativistic heavy-ion collisions at the LHC}

\date{\today  \hspace{1ex}}
\author{Jun Yan}
\affiliation{Key Laboratory of Quark \& Lepton Physics (MOE) and Institute of Particle Physics,
 Central China Normal University, Wuhan 430079, China}

\author{Shi-Yong Chen \footnote{chensy@mails.ccnu.edu.cn}}
\affiliation{Key Laboratory of Quark \& Lepton Physics (MOE) and Institute of Particle Physics,
 Central China Normal University, Wuhan 430079, China}

\author{Wei Dai}
\affiliation{School of Mathematics and Physics, China University of Geosciences, Wuhan 430074, China}

\author{Ben-Wei Zhang\footnote{bwzhang@mail.ccnu.edu.cn}}
\affiliation{Key Laboratory of Quark \& Lepton Physics (MOE) and Institute of Particle Physics,
 Central China Normal University, Wuhan 430079, China}
 \affiliation{Institute of Quantum Matter, South China Normal University, Guangzhou 510006, China}

\author{Enke Wang}
\affiliation{Institute of Quantum Matter, South China Normal University, Guangzhou 510006, China}
\affiliation{Key Laboratory of Quark \& Lepton Physics (MOE) and Institute of Particle Physics,
 Central China Normal University, Wuhan 430079, China}

\begin{abstract}
In this paper we investigate the medium modifications of girth distributions for inclusive jets and $Z^0$ tagged jets
with small radius ($R=0.2$) in  Pb+Pb collisions with $\sqrt{s}=2.76$~TeV at the LHC. The partonic spectrum in the initial hard scattering of elementary collisions are obtained by a event generator POWHEG+PYTHIA, which matches the next-to-leading (NLO) matrix elements with parton showering, and energy loss of fast parton traversing in hot/dense QCD medium is calculated by Monte Carlo simulation within Higher-Twist formalism of jet quenching in heavy-ion collisions. We present the model calculations of event normalized girth
distributions for inclusive jets in p+p and Pb+Pb collisions at $\sqrt{s}=2.76$~TeV, which give nice descriptions of ALICE measurements.  It is shown that the girth distributions of inclusive jets in Pb+Pb are shifted to lower girth region relative to that
in p+p.  Thus the nuclear modification factor of girth distributions for inclusive jets is larger than unity at small girth region, while smaller than one at large girth region. This behavior results from more soft fragments inside a jet as well as the fraction alteration of gluon/quark initiated jets in heavy-ion collisions.   We further predict the girth distributions for $Z^0$ boson tagged jets in Pb+Pb collisions at $\sqrt{s}=2.76$~TeV, and demonstrate that the medium modification on girth distributions for $Z^0$ tagged jets is less pronounced as compared to that for inclusive jets because the dominant components of $Z^0$ tagged jets are quark-initiated jets.

\end{abstract}


\maketitle

\section{Introduction}
\label{sec:Intro}

Energetic partons produced at early stage of relativistic heavy-ion collisions (HIC) may lose substantial energy through interacting with the quark-gluon plasma (QGP), a novel nuclear medium under extreme conditions of high temperature and energy density to be formed in these collisions. This
phenomenon is known as jet quenching~\cite{Wang:1991xy,Gyulassy:2003mc,Qin:2015srf}, which could provide promising tools to investigate the creation and properties of the QGP.  In the last decade the experimental and theoretical studies of jet quenching have been extended from the suppression of leading hadron productions~\cite{Khachatryan:2016odn, Acharya:2018qsh, Aad:2015wga, Burke:2013yra,Chen:2010te,Chen:2011vt,Liu:2015vna,Dai:2015dxa,Dai:2017piq,Dai:2017tuy,Ma:2018swx,Xie:2019oxg} to medium modifications of a wealth of full jet observables, such as inclusive jets, di-jets, gauge bosons tagged jets, heavy flavor jets as well as jet substructures~\cite{Vitev:2008rz,Vitev:2009rd,Aad:2010bu,Chatrchyan:2011sx,Chatrchyan:2012gt,Aad:2014bxa,Chatrchyan:2012gw,Chatrchyan:2013kwa, Aad:2014wha,Sirunyan:2017jic, CasalderreySolana:2010eh, Sirunyan:2018ncy,Young:2011qx,He:2011pd,ColemanSmith:2012vr,Neufeld:2010fj,Zapp:2012ak, Dai:2012am, Ma:2013pha, Senzel:2013dta, Casalderrey-Solana:2014bpa,Milhano:2015mng,Chang:2016gjp,Majumder:2014gda, Chen:2016cof, Chien:2016led, Apolinario:2017qay,Connors:2017ptx,Zhang:2018urd,Dai:2018mhw,Luo:2018pto,Chang:2019sae,Wang:2019xey,Chen:2019gqo,Chen:2020kex}.
A full jet is a collimated spray of final-state hadrons in  $e^+e^-$ collisions, elementary hadron-hadron collisions and the nucleus-nucleus reactions with a large center-of-mass colliding energies, and the existence of the QGP should naturally alter the yields and the internal structures of a full jet, and thus the medium modifications of jet observables could be used for tomography of the nuclear matter formed in HIC.

One interesting jet substructure observable is jet girth (or angularity), which probes the radiation inside a jet~\cite{Giele:1997hd,Acharya:2018uvf,KunnawalkamElayavalli:2017hxo,Agafonova:2019tqe,Wan:2018zpq}.  The medium modification of jet girth may help us understand deeper the specific features of jet-medium interaction by providing complementary aspects of the jet fragmentation such as jet transverse momentum distribution, and shed light on how the jet substructure is resolved by the medium in HIC. Recently, ALICE Collaboration has measured the normalized girth distributions for an inclusive jets with small-radius  ($R=0.2$) in heavy-ion collisions~\cite{Acharya:2018uvf}, which further facilitates the studies of girth in HIC since the theoretical model calculations now could be confronted directly with the data to infer some crucial information of jet propagation in the QCD medium.

In this work, we present our study on the normalized distributions of girth for inclusive jets and $Z^0$ tagged jets with jet radius $R=0.2$ both in Pb+Pb collisions at $\sqrt{s}=2.76$~TeV. We utilize POWHEG+PYTHIA~\cite{Alioli:2010xa,Alioli:2010qp,Buckley:2016bhy},  a model matching next-to-leading order (NLO) with parton shower (PS) including hadronization in the final state to obtain the solid baseline calculations of jet girth in p+p collisions, which are then combined with a numerical simulation of parton energy loss within the higher-twist approach of jet quenching~\cite{Guo:2000nz,Zhang:2003yn,Zhang:2003wk,Majumder:2009ge} to compute the girth distribution in high-energy nuclear collisions. Our numerical results of girth distribution for inclusive jets give satisfactory descriptions of ALICE data both in p+p and Pb+Pb collisions, where we observe a shift of girth distribution to lower values in Pb+Pb relative to that in p+p. The girth distributions of $Z^0+{\rm jet}$ in Pb+Pb collisions at the LHC are calculated for the first time, which show medium modifications of girth distributions for $Z^0+{\rm jet}$ are less pronounced than that for inclusive jets.

This paper is organized as follows.  In Sec.~\ref{sec:framework} we will introduce the framework of computing the normalized distributions of girth in detail in p+p and Pb+Pb collisions.  The numerical results and detailed discussions of the medium modifications of girth in inclusive jets and gauge boson $Z^0$ tagged jets are presented in Section~\ref{sec:results}. And Sec.~\ref{sec:summary} summarizes our study.

\section{analysis framework}
\label{sec:framework}

We study a jet substructure observable, the girth, which characterizes the radial distribution of radiation inside a jet~\cite{Giele:1997hd,Acharya:2018uvf}, and is defined
as:
\begin{eqnarray}
g=\sum_{i}\frac{p_{T,i}}{p_{T,{\rm jet}}}\left| \Delta R_{i,{\rm jet}}\right|
\label{eq:g}
\end{eqnarray}
Where $p_{T,i}$ stands for the transverse momentum of $i$th constituent inside the jet with transverse momentum $p_{T,{\rm jet}}$. $\Delta R_(i,jet)$ denotes the distance in $(\eta,\phi)$ space between this constituent and the jet axis. Girth is sensitive to the radial energy profile of the jet, and for fixed jet $p_T$, the jet is more collimated with a lower value of girth.

In this paper a Monte Carlo model POWHEG+PYTHIA, with next-to-leading order (NLO) matrix elements matched with parton showering~\cite{Alioli:2010xa,Alioli:2010qp,Buckley:2016bhy}, is used to simulate hadron productions in p+p collisions. In our simulation the POWHEG BOX code is utilized~\cite{Alioli:2010xa,Alioli:2010qp}, which provides a computer framework for implementing NLO calculations in parton  shower Monte Carlo programs according to the POWHEG scheme~\cite{Frixione:2007vw}. It has been shown that, POWHEG BOX Monte Carlo program matched with parton showering can make a good description of productions and correlations for a variety of processes, such as single-top production, di-jet, gauge boson+jets, multiple jets etc.~\cite{powheg-box}. We generate the NLO matrix elements for QCD dijet processes with POWHEG BOX, and then matched with PYTHIA to simulate parton showering and hadronization~\cite{Sjostrand:2006za}. Fastjet package~\cite{Cacciari:2008gp} is used to reconstruct final state hadrons into full jets.

\begin{figure}[!htb]
\centering
\includegraphics[width=9.5cm,height=9cm]{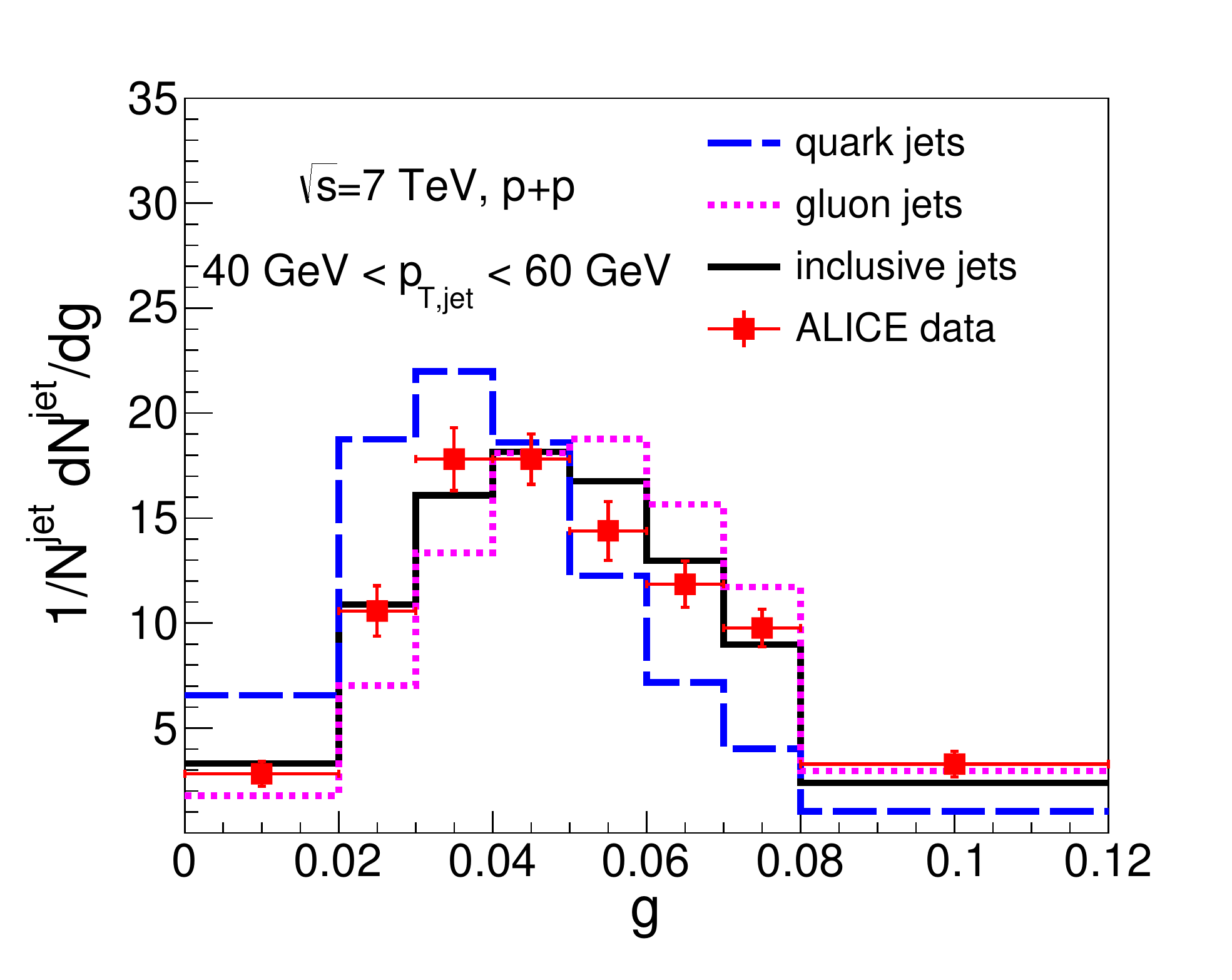} \\
\caption{Normalized girth distribution for inclusive jets in p+p collisions at
$\sqrt{s}=7$~TeV from POWHEG+PYTHIA calculation as compared with ALICE data~\cite{Acharya:2018uvf}.}
\label{g-ALICE}
\end{figure}

To compare with experimental data,
events are selected within the same kinematic
cuts as imposed in the experimental measurements that
we will compare to. In ALICE Collaboration data, jets are reconstructed by the anti-$k_{T}$
algorithm with small radius $R=0.2$ from charged tracks where a cutoff with $p_{T}>0.15$~GeV is imposed.
The transverse momentum of jets are required to $40 \ {\rm GeV} <p_{T, \rm jet}<60 $~GeV in central rapidity region $\left|\eta_{\rm jet}\right|<0.7$. Our simulation results of
normalized distributions of girth in p+p collisions at
$\sqrt{s}=7$~TeV and their comparison with ALICE data are plotted in Fig.~\ref{g-ALICE}. One can observe that, POWHEG+PYTHIA calculation can provide reasonable description for girth distributions in p+p collisions in the overall region, which will be served as baseline for the subsequent study of nuclear modification in heavy-ion collisions. We also plot the girth distributions of quark-initiated and gluon-initiated jets respectively at $\sqrt{s}=7$~TeV in Fig.~\ref{g-ALICE}. We observe that at the same jet $p_T$, girth distributions of gluon jets are broader than quark jets, and the peak of girth distribution for gluon jets is located at larger $g$ relative to that for quark jets. It implies as compared to quark jets, gluon jets have larger radiation width, and favor harder radiation with wider radiation angle, on average.

In heavy-ion collisions, partons produced from hard scattering will lose their energy due to jet-medium interactions. In our analysis framework, with the initial jet shower partons generated by POWHEG+PYTHIA and the initial position of these partons in QGP sampled from Glauber model~\cite{Alver:2008aq}, they propagate through the quark-gluon plasma step-by-step after the formation of QGP. The probability for gluon
radiation happens in QGP during a time step $\Delta t$ can be expressed as~\cite{He:2015pra,Cao:2016gvr,Cao:2017hhk,Wang:2019xey}:
\begin{eqnarray}
P_{rad}(t,\Delta t)=1-e^{-\left\langle N(t,\Delta t)\right\rangle} \, .
\label{eq:g}
\end{eqnarray}

Here $\left\langle N(t,\Delta t)\right\rangle$ is the average number of emitted gluons integrated
from the medium induced radiated gluon spectrum within Higher-Twist(HT) approach~\cite{Guo:2000nz,Zhang:2003yn,Zhang:2003wk,Majumder:2009ge}:
\begin{eqnarray}
\frac{dN}{dxdk^{2}_{\perp}dt}=\frac{2\alpha_{s}C_sP(x)\hat{q}}{\pi k^{4}_{\perp}}\sin^2(\frac{t-t_i}{2\tau_f})(\frac{k^2_{\perp}}{k^2_{\perp}+x^2M^2})^4
\label{eq:g}
\end{eqnarray}

Here $\alpha_{s}$ is the strong coupling constant, $M$ the mass of parent parton, $x$ denotes the energy fraction relative to mother parton, and $k_\perp$  gives transverse momentum of the radiated gluon.  We have applied a lower energy cut-off for the
emitted gluon $x_{min}=\mu_{D}/E$ in the calculation, with $\mu_{D}$ the Debye
screening mass. $C_s$ is the Casimir factor for quarks ($C_F$) and gluons ($C_A$), and $P(x)$ stands for the splitting function in vacuum, $\tau_f=2Ex(1-x)/(k^2_\perp+x^2M^2)$ is the formation time of the radiated gluons.  The jet transport parameter $\hat{q}$ is proportional to the local parton density distribution in the QCD medium, and related to the space and time evolution of the medium relative to its initial value $\hat{q}_0$ in the central region when QGP formed, which controls the magnitude of energy loss due to jet-medium interaction.

Multiple gluon radiation is allowed during each time step, where the number of
radiated gluons is assumed to follow a Poisson distribution.
The values of $x$ and $k_\perp$ could be sampled according to the radiative gluon spectrum in Eq.(\ref{eq:g}). A Hard Thermal Loop (HTL) approximation formula~\cite{Neufeld:2010xi} $\frac{dE^{coll}}{dt}=\frac{\alpha_{s}C_{s}\mu^{2}_{D}}{2}ln\frac{\sqrt{ET}}{\mu_{D}}$ has been adopted to simulate the collisional
energy loss of the showered partons~\cite{Dai:2018mhw,Wang:2019xey}.
The evolution profile of the QGP medium is provided by
the smooth iEBE-VISHNU hydro model~\cite{Shen:2014vra}. In our simulation, $\hat{q}_{0}=1.2$~GeV$^2$/fm is directly extracted from the sophisticated study of the identified hadron suppression in Pb+Pb collisions at $2.76$~TeV using the same evolved QGP medium expansion~\cite{Ma:2018swx}. Therefore, an effective jet transport parameter and $\hat{q}$ which only consider jet transport in QGP phase is used.
Jet partons stop their propagation on the freeze-out hypersurface
of the fireball ($T_c=165$~MeV).
After all the partons propagate through QGP, PYQUEN method is used to preform the hadronization process~\cite{Lokhtin:2000wm,Lokhtin:2005px}. In the model, the radiated gluons are rearranged in the same string as their parent partons,
and these partons could fragment into hadrons
by standard PYTHIA hadronization procedure.

\section{Results and discussion}
\label{sec:results}

\begin{figure}[!htb]
\centering
\includegraphics[width=9.5cm,height=9cm]{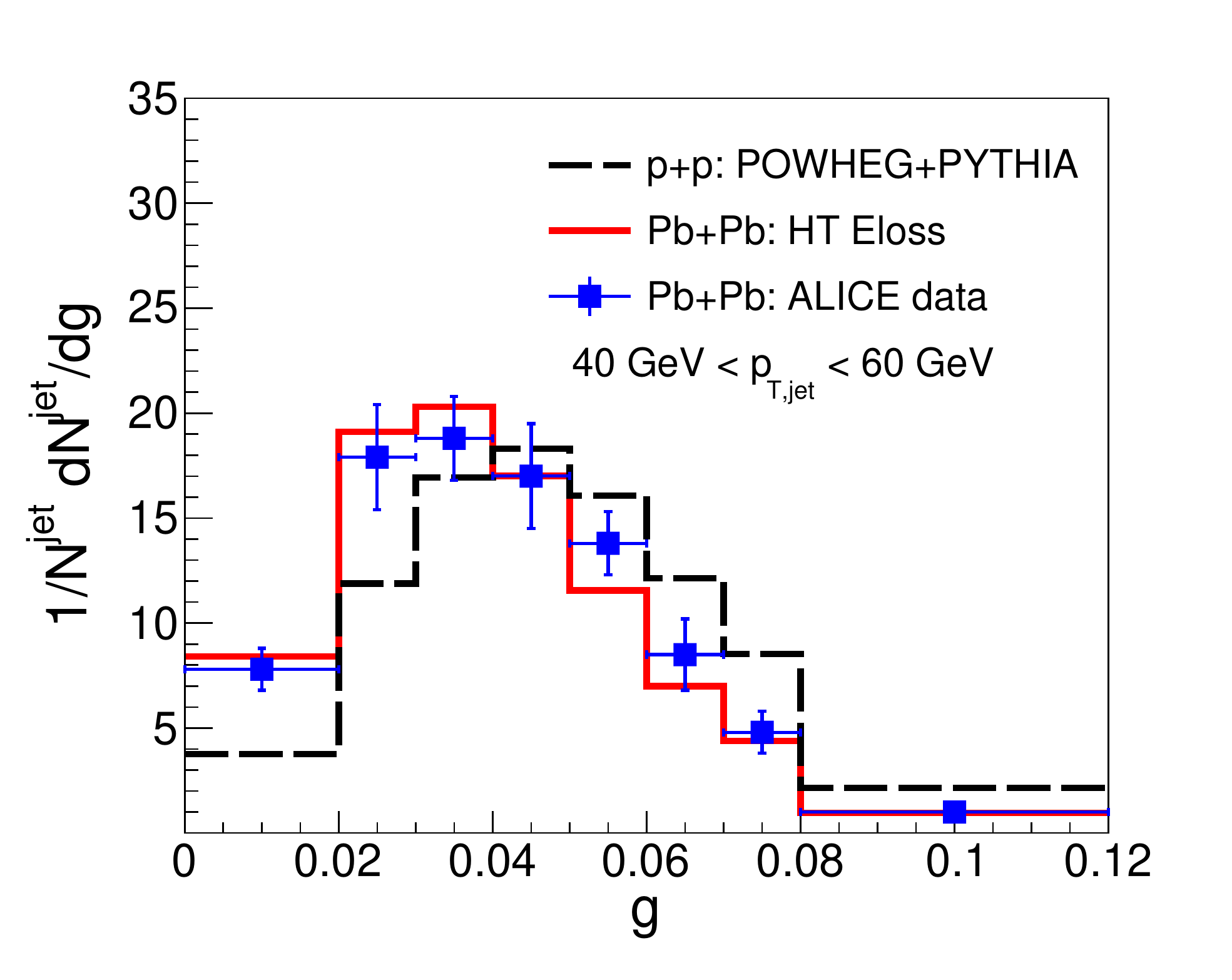} \\
\caption{Normalized girth distributions of inclusive jets in p+p and Pb+Pb collisions at $\sqrt{s}=2.76$~TeV as compared with ALICE data~\cite{Acharya:2018uvf}.}
\label{g-AA}
\end{figure}

We now could calculate
the jet number normalized girth distributions in Pb+Pb collisions at $\sqrt{s}=2.76$~TeV.
We apply the same kinematic cuts as we did in p+p collisions in the previous section.
Shown in Fig.~\ref{g-AA} are our numerical results for girth distributions for inclusive jets with radius $R=0.2$, which are  confronted  against ALICE measurements~\cite{Acharya:2018uvf}. Satisfactory agreement can be found between our theoretical calculations and experimental data. Compared with that in p+p collisions, the observed girth distribution in Pb+Pb is shifted to lower values of $g$ due to jet-medium interactions, which implies that  a jet in Pb+Pb may exhibit smaller radiation width and has more softer fragments than that in p+p collisions.

To compare the difference of girth distributions in HIC and those in p+p in a more straightforward way, we can plot numerically
the nuclear modification ratio of girth distributions $R_{AA}^{\rm girth}$, which is defined as:
\begin{eqnarray}
R_{AA}^{\rm girth}={\dfrac{1}{N_{AA}} \dfrac{dN_{AA}}{dg}}/{\dfrac{1}{N_{pp}} \dfrac{dN_{pp}}{dg}} \,\, .
\label{eq:ratio}
\end{eqnarray}

\begin{figure}[!htb]
\centering
\includegraphics[width=9.5cm,height=9cm]{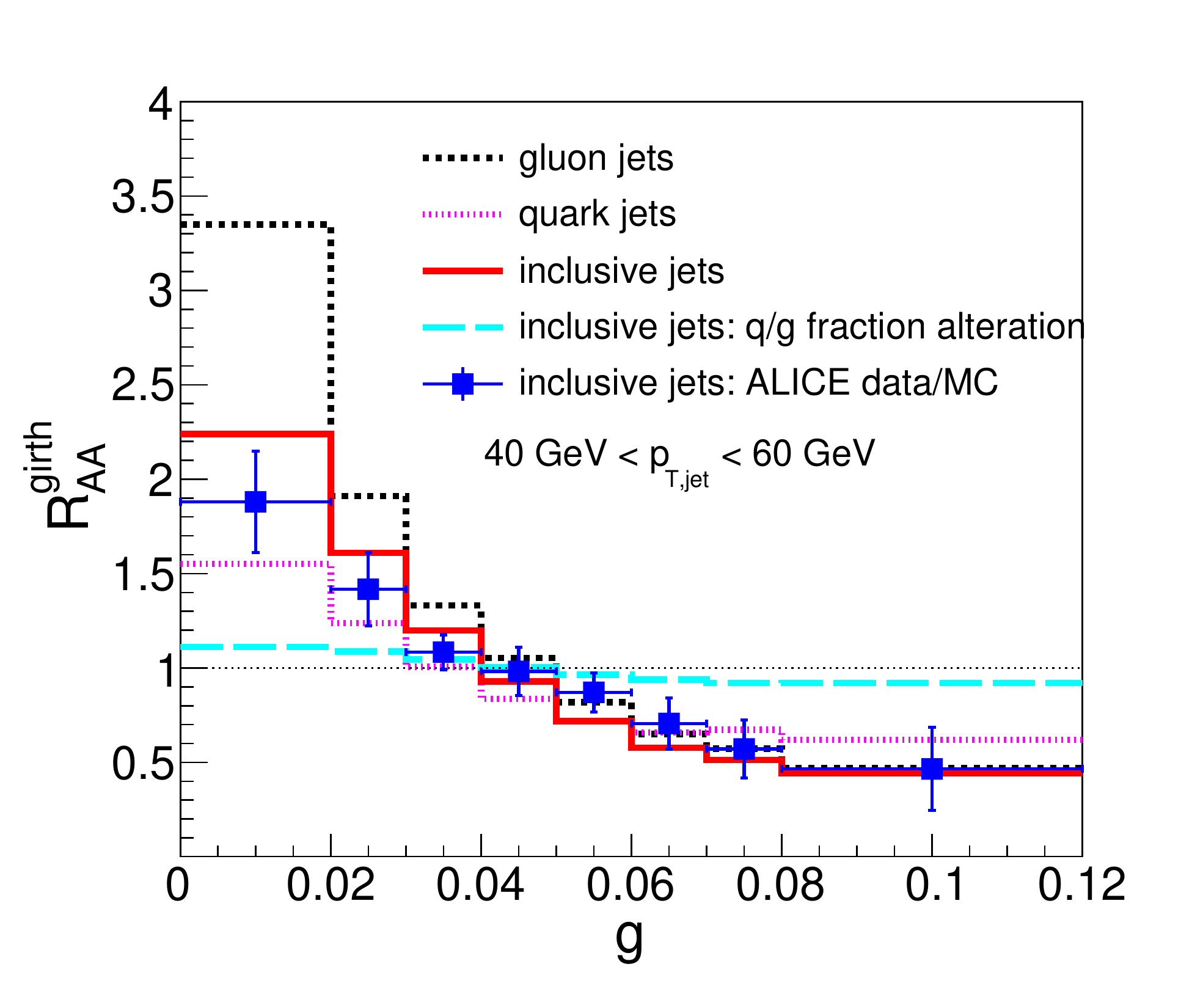} \\
\caption{Nuclear modification ratio of girth distribution of inclusive jets as well as quark and gluon jets. The ratio from ALICE Collaboration is preformed by Pb+Pb measurements scaled by MC simulation in p+p~\cite{Acharya:2018uvf}.
}
\label{ratio-jets}
\end{figure}

Present in Fig.~\ref{ratio-jets} are $R_{AA}^{\rm girth}$ of jet number normalized girth distributions for inclusive jets, as well as the components of inclusive jets: quark jets and gluon jets. One can see  there is an enhancement of girth distribution for both quark jets and gluon jets at small $g$ region, whereas a suppression at large $g$ region. And the nuclear modification for gluon jets girth distribution is stronger than that for quark jets.  Because inclusive jets consist of both gluon jets and quark jets, it is natural to observe that the curve of $R_{AA}^{\rm girth}$ for inclusive jets goes between the curve of $R_{AA}^{\rm girth}$ for gluon jets and the one for quark jets. The overall shifting of girth distribution for gluon jets and quark jets results from the more soft hadron fragments in these jets in HIC, as we may demonstrate in detail later.

Another reason underlying the nuclear modification of girth distribution for inclusive jets is the change of relative fraction of quark and gluon jets. In higher-twist formalism of parton energy loss (as most of other models of jet quenching), gluons may suffer more energy loss than quarks in QCD medium with its larger color charge. So generally we should see the relative fraction of quark jets should increase in A+A relative to p+p collisions, and because girth distribution of quark jets is spread over smaller $g$ region as compared to that of gluon jets, as illustrated in Fig.~\ref{g-ALICE}, the solo impact of fraction changes of quark jets and gluon jets to form inclusive jets will increase girth distribution of inclusive jets at small $g$ and decrease it at large $g$. This is demonstrated in Fig.~\ref{ratio-jets}, where the curve labelled ‘inclusive jets: q/g fraction alteration’ represents the numerical result of $R_{AA}^{\rm girth}$ by only considering the effect of quark/gluon jets fraction changes due to energy loss while assuming there are no medium modifications for girth distributions of pure gluon jets and quark jets in heavy-ion collisions.

\begin{figure}[!htb]
\centering
\includegraphics[width=9.5cm,height=9cm]{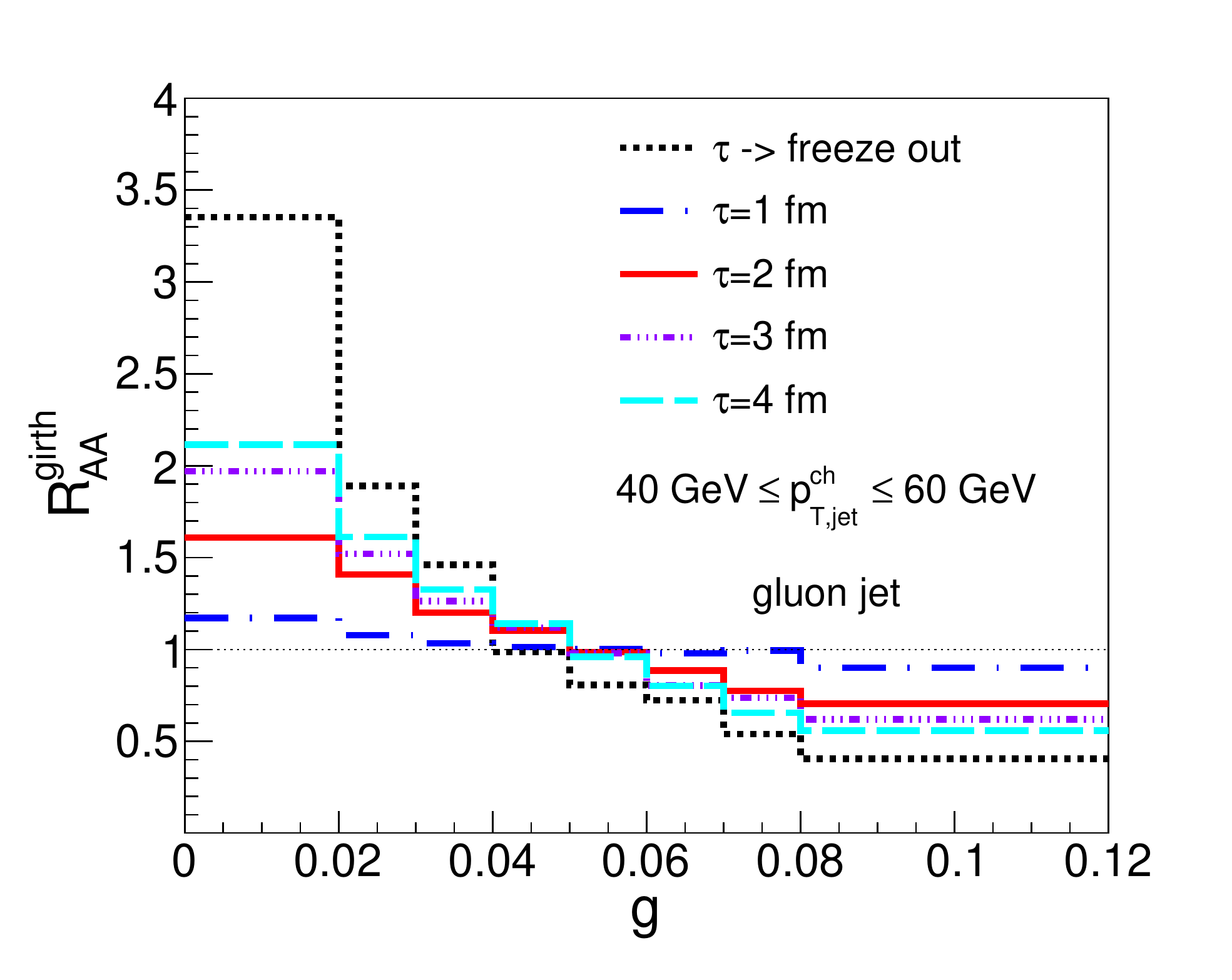}
\caption{Extended jet transverse profile of quark(top panel) and gluon(bottom panel) jets in p+p collisions at $\sqrt{s}=2.76$~TeV.}
\label{time}
\end{figure}

To see the time evolution of jet girth distribution in QGP, we plot $R_{AA}^{girth}$ for gluon jets as an example at time steps $\tau=1,2,3,4$~GeV and their comparison with full evolution in Fig.~\ref{time}. One can observe that when jet suffer energy loss, there is an enhancement of girth distribution in small $g$ region, while a suppression at large $g$ region.
The muclear modifications will be more visiable as jet propagate in QGP gradually.

\begin{figure}[!htb]
\centering
\includegraphics[width=8cm,height=7cm]{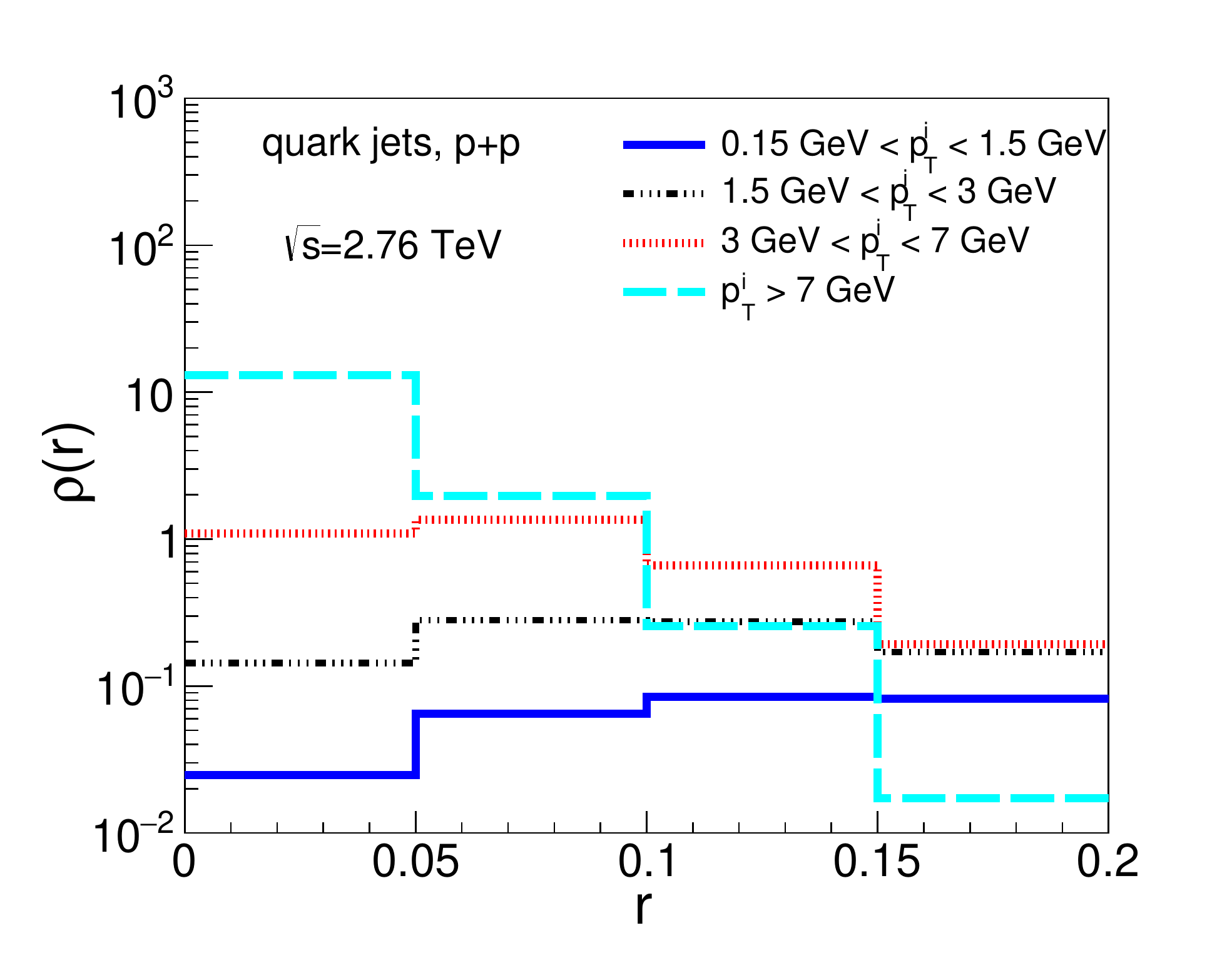}
\includegraphics[width=8cm,height=7cm]{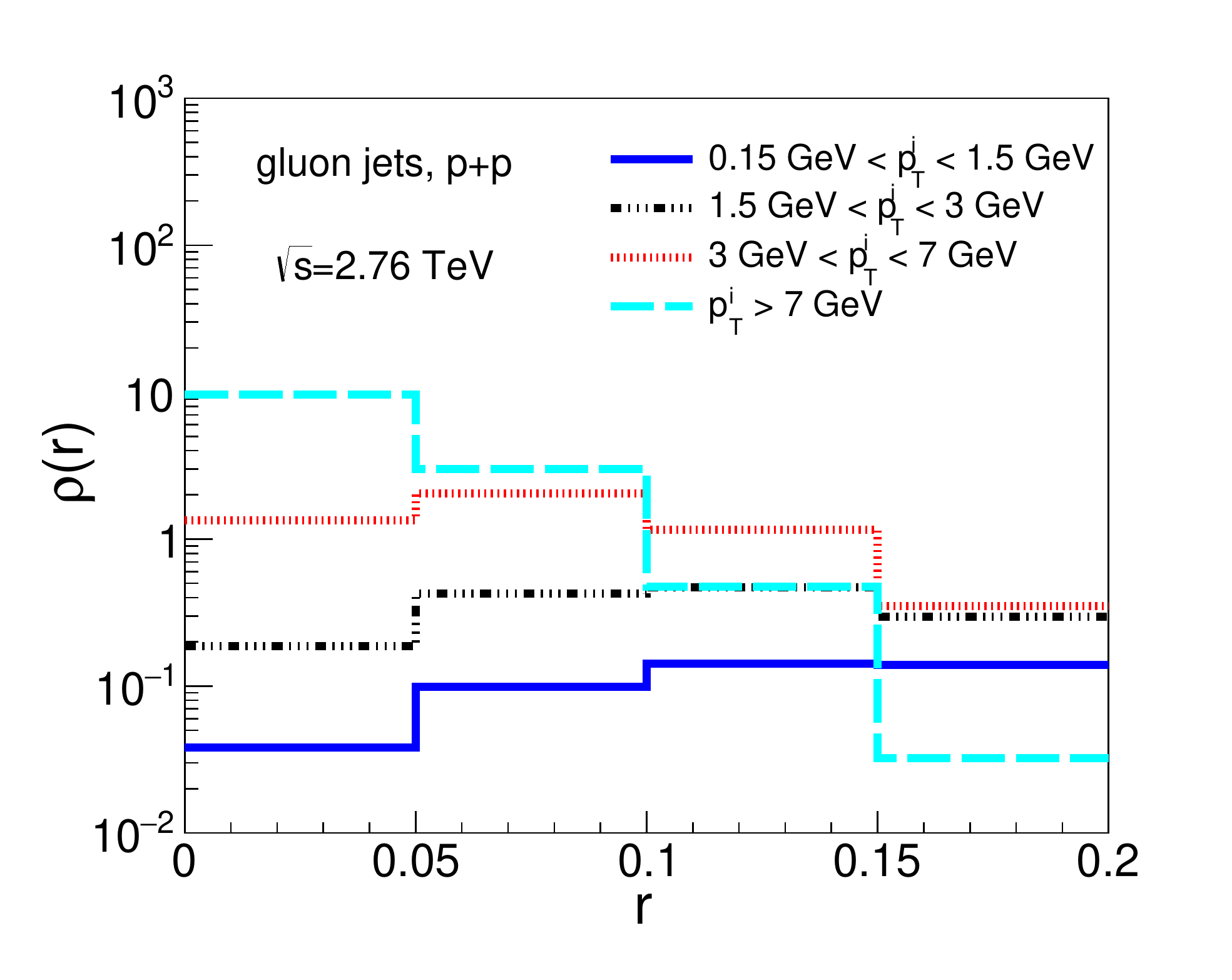} \\
\caption{Extended jet transverse profile of quark(top panel) and gluon(bottom panel) jets in p+p collisions at $\sqrt{s}=2.76$~TeV.}
\label{shape-pp}
\end{figure}

To understand why for normalized quark (gluon) jet girth distributions $R_{AA}^{\rm girth} >1$ at small $g$ region, and $R_{AA}^{\rm girth} <1$ at large $g$ region, as shown in Fig.~\ref{ratio-jets}, we turn to another very closely related jet substructure observable of girth, jet shape (or jet profile)~\cite{Vitev:2008rz,Vitev:2009rd,Chatrchyan:2013kwa,Sirunyan:2018ncy}. Actually, girth gives the first radial moment of
jet shape~\cite{Giele:1997hd}. Jet shape characterizes the averaged transverse momentum profile of jet in $\Delta r$ which is defined as:
\begin{eqnarray}
\rho(r)=\frac{1}{\Delta{r}}\frac{1}{N_{\rm jet}}\sum_{\rm jet}\frac{\sum_{i}p_{T,\rm jet}^{i}(r-\Delta{r}/2,r+\Delta{r}/2)}{p_{T,\rm jet}(0,R)}
\label{eq:rho}
\end{eqnarray}
where $r$ is a distance to jet axis inside jet. $p_{T,\rm jet}^{i}$ is the transverse momentum of $i$th constituents inside jets. It is noted that, in our model all the showered partons should lose energy when they transverse QGP due to jet quenching effect. After they transverse through QGP and hadronization, the hadrons fragmented from the less energy parton may fall off the kinematic cut of jet candidate. On the other hand, medium induced gluon radiation will provide additional particles in the event list.

\begin{figure}[!htb]
\centering
\includegraphics[width=8cm,height=7cm]{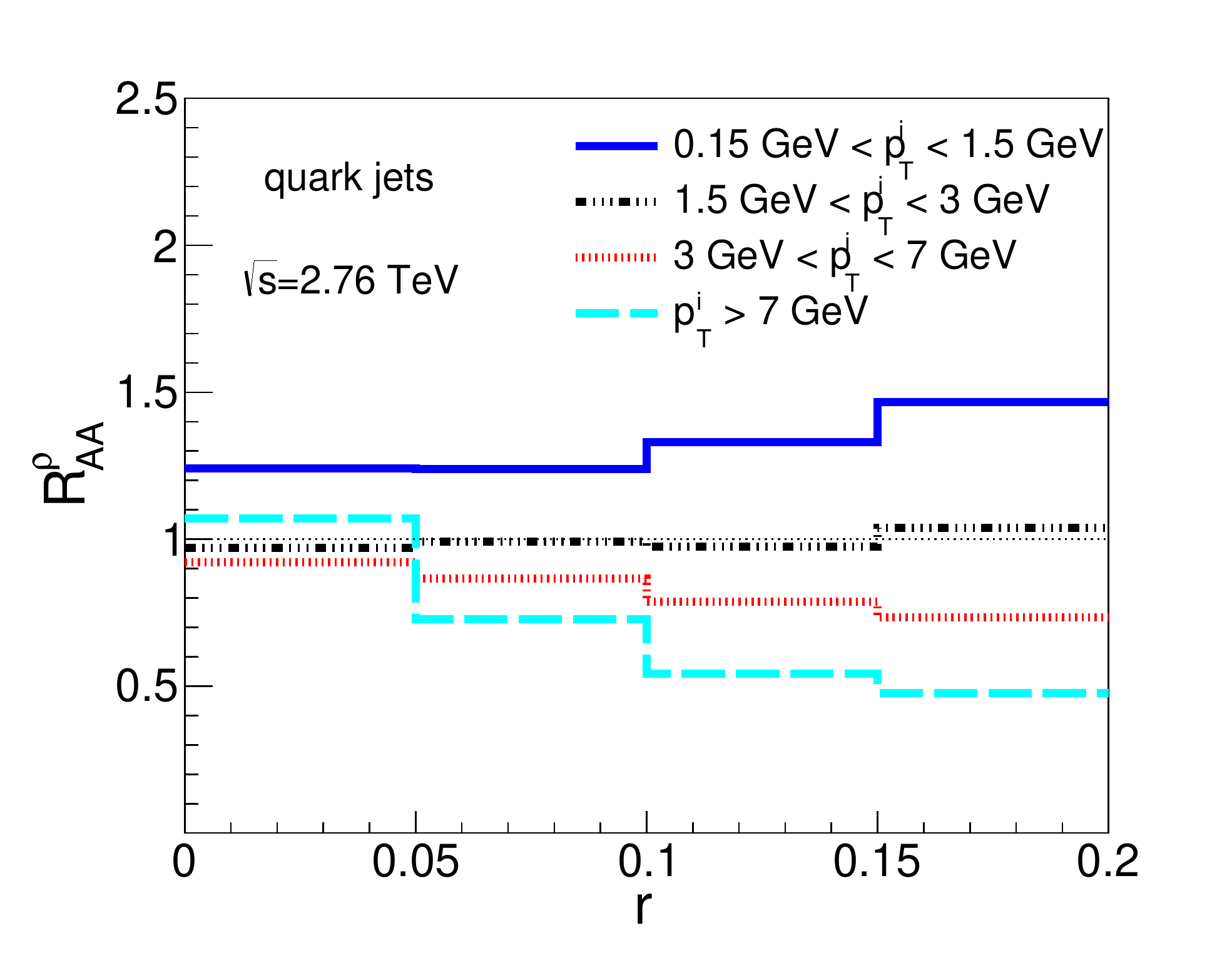}
\includegraphics[width=8cm,height=7cm]{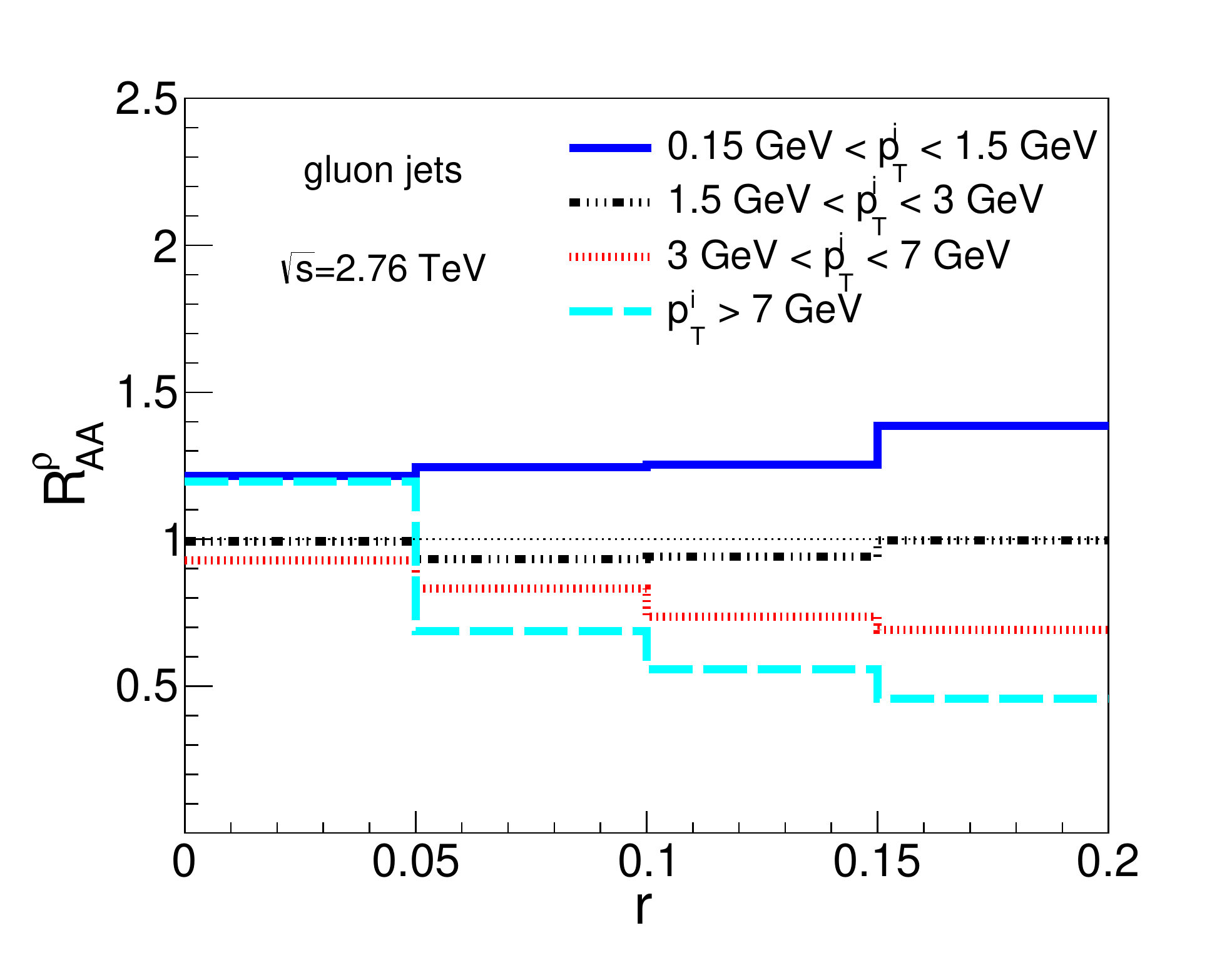} \\
\caption{Nuclear modification ratio of extended jet transverse profile of quark(top panel) and gluon(bottom panel) jets in Pb+Pb to that in p+p collisions at $\sqrt{s}=2.76$~TeV.}
\label{shape-ratio}
\end{figure}

Fig.~\ref{shape-pp} demonstrate the transverse momentum distributions of quark and gluon jets inside jet cone calculated from jet constituents by four $p_{T}$ bins. We found that the constituents with highest $p_{T}$ are more likely distributed closely to jet axis for both quark and gluon jets. We plot the nuclear modification factor of jet shape $R_{AA}^{\rho}=\rho_{AA}/\rho_{pp}$ in Fig.~\ref{shape-ratio}. It can be easily observed that distributions of particles with low $p_T$ are enhanced in Pb+Pb relative to in p+p, whereas distributions of particles with very large $p_T$ are suppressed, which holds true for both gluon and quark jets. The enhanced distribution of soft fragments inside a jet can lead to $R_{AA}^{\rm girth} >1 $ at small girth for both gluon and quark jets.

To see the impact of different medium modifications of girth distributions for gluon and quark jets, we also
calculate the event normalized distribution of girth of  $Z^0$ tagged jets which are dominated by quark jets both in p+p and Pb+Pb collisions. We adopt the same kinematic cuts by CMS experiment~\cite{Sirunyan:2017jic} to select $Z^0$ tagged jets events in both p+p and Pb+Pb collisions.
Two decay channels are taken into account during the reconstruction of the massive gauge boson $Z^0$: $Z^0\rightarrow e^+e^-$ and $Z^0 \rightarrow \mu^+\mu^-$.
We choose events that electrons have $p_{T}^{e}>20 $ GeV, $|\eta^{e}|<2.5$, and muons have $p_{T}^{\mu}>10 $ GeV, $|\eta^{\mu}|<2.4$.
Then the $Z^0$ bosons are reconstructed by lepton pairs with mass $70 \ {\rm GeV}<M_{ll}<110$~GeV and $p_{T}^{Z}>40$ GeV.
Jets are constructed by FASTJET from final-state charged hadrons with the anti-$k_T$ algorithm and jet radius $R=0.2$. All the jets tagged by $Z^0$ are required to pass the threshold $ p_{T,\rm jet}>30$ GeV, and are rejected in the radius of $R <0.4 $ from a lepton to suppress jet energy contamination. Jets are further chosen to have $40 \ {\rm GeV} <p_{T, \rm jet}<60 $~GeV and $\left| \eta_{\rm jet} \right|<0.7$.

\begin{figure}[!htb]
\centering
\includegraphics[width=9.5cm,height=8cm]{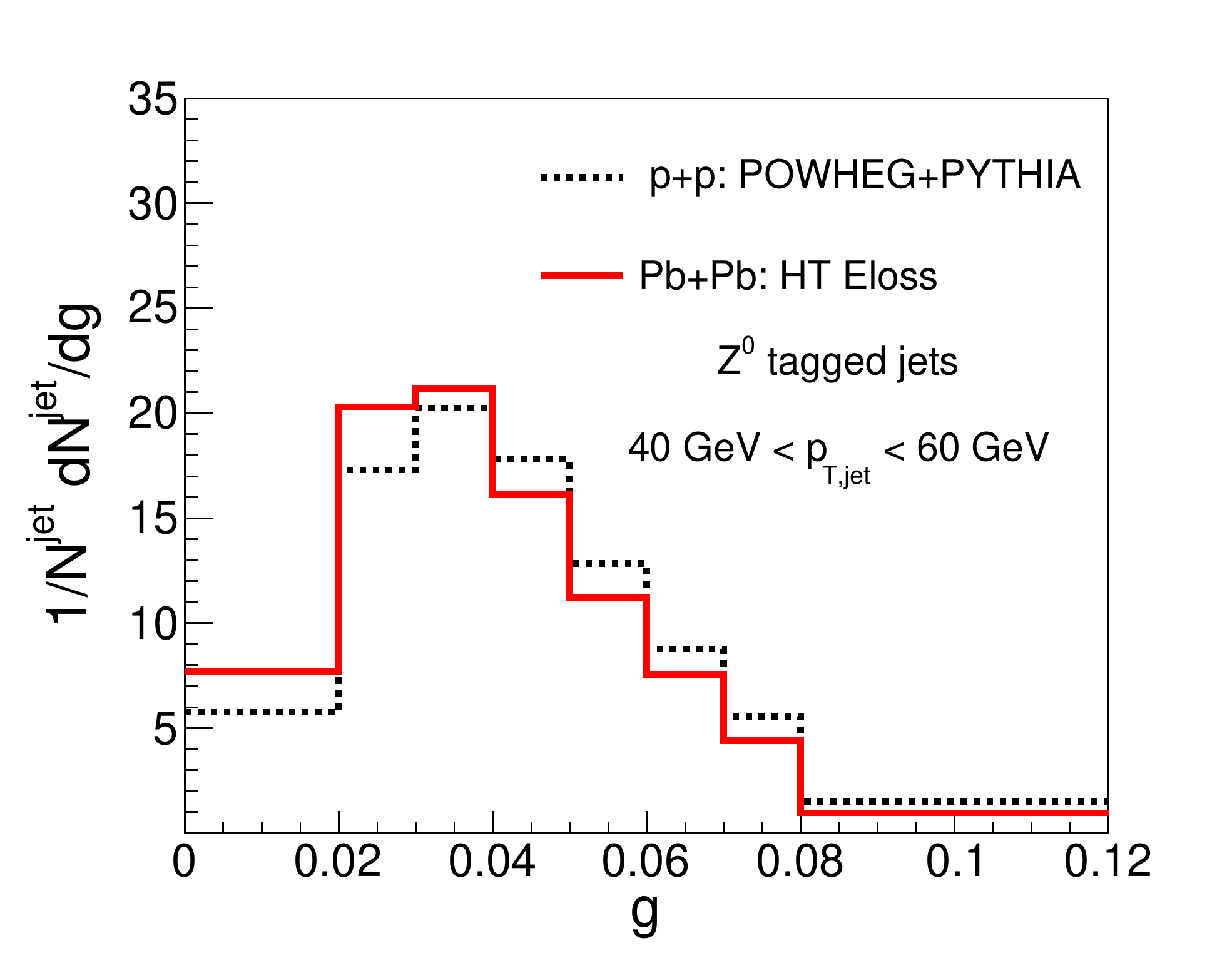} \\
\caption{Normalized girth distributions of $Z^0$ tagged jets in p+p and Pb+Pb collisions at $\sqrt{s}=2.76$~TeV.}
\label{zjet-girth}
\end{figure}

\begin{figure}[!htb]
\centering
\includegraphics[width=9.5cm,height=8cm]{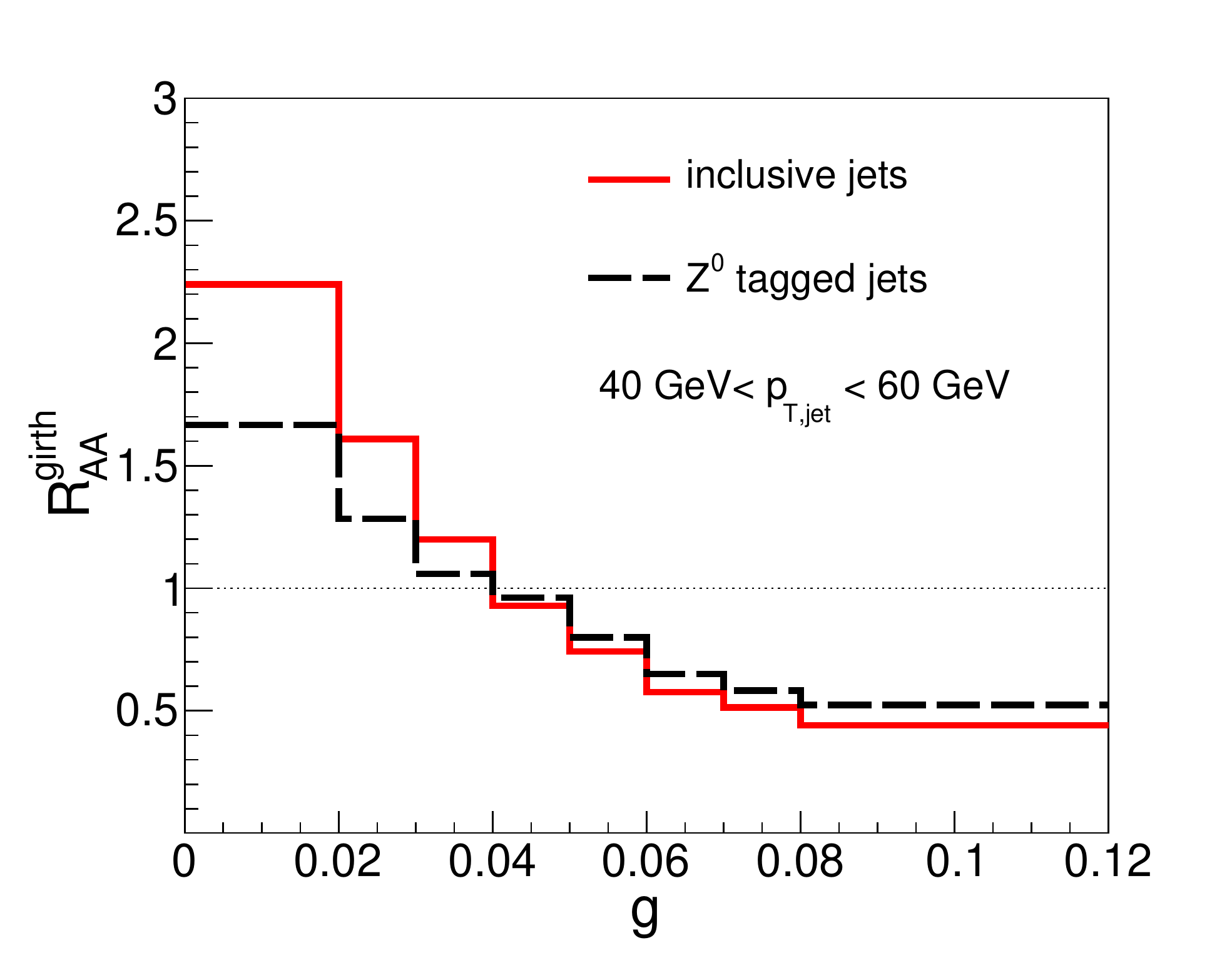} \\
\caption{Nuclear modification ratios of girth distributions for inclusive jets and $Z^0$ tagged jets in Pb+Pb collisions at $\sqrt{s}=2.76$~TeV.}
\label{ratio-zjets}
\end{figure}

The normalized distributions of girth for $Z^0$ tagged jets in p+p and Pb+Pb collisions at $\sqrt{s}=2.76$~TeV are plotted in Fig.~\ref{zjet-girth}. Compared to that in p+p collisions, the girth distributions in Pb+Pb collisions are also shifted to lower $g$ region, which is the same as we have observed for girth distributions for inclusive jets. We further calculate the nuclear modification ratio of girth distributions for $Z^0$ tagged jets, as shown in Fig.~\ref{ratio-zjets}.
The nuclear modifications of girth distributions for inclusive jets are much stronger than that for $Z^0$ tagged jets. This distinction is to a large extent attributed to the flavor composition of $Z^0$ tagged jets and inclusive jets.
As shown in Table.~\ref{frac} $Z^0$ tagged jets are dominated by quark jets while the inclusive jets have a more significant fraction of gluon jets. Because nuclear modification of girth distribution for gluon jets are larger than that for quark jets (see for example Fig.~\ref{ratio-jets}), the nuclear modification of girth distributions for $Z^0$ tagged jets should be weaker as compared with that for inclusive jets.


\begin{table}[!htb]
\begin{center}
\begin{tabular}{|c|c|c|}
	\hline quark jet fraction & p+p & Pb+Pb  \\
	\hline inclusive jets & 33\% & 45\% \\
	\hline $Z^0$ tagged jets & 88\% & 92\% \\
	\hline
\end{tabular}
\caption{Estimation of quark jet fraction of inclusive jets and $Z^0$ tagged jets in p+p and Pb+Pb collisions at leading-order (LO), with jet radius $R=0.2$, $40 \ {\rm GeV} < p_{T,\rm jet} < 60$~GeV.}
\label{frac}
\end{center}
\end{table}


\section{Summary}
\label{sec:summary}
In this paper, with a NLO$+$PS event generator POWHEG$+$PYTHIA for p+p baseline and HT parton energy loss formalism for jet quenching, we have studied
the nuclear modifications of girth distributions for both inclusive jets and $Z^0$ tagged jets with small radius $R=0.2$ in Pb+Pb at $\sqrt{s}=2.76$~TeV.
Our numerical results of inclusive jets could provide reasonable description of ALICE data. We show girth distributions for inclusive jets
are shifted to lower girth region in Pb+Pb collisions compared to that in p+p collisions, and similar trends have also been observed for quark and gluon jets.
We find two factors contributing to the nuclear modifications of girth distributions: more soft fragments inside a jet (inclusive jets, quark jets and gluon jets) after jet-medium interaction, and the enhanced fraction of quark-initiated jets in HIC.
Girth distributions for gluon jets show stronger nuclear modifications than that for quark jets in HIC, since gluons may suffer more energy loss than quarks in our model.
We further give the prediction of girth distributions for $Z^0$ boson tagged jets in both p+p and Pb+Pb collisions at $\sqrt{s}=2.76$~TeV.
The medium modification on girth distributions for $Z^0$ tagged jets is found to be less pronounced as compared to that for inclusive jets,
since $Z^0$ tagged jets are dominant by quark-initiated jets.

{\bf Acknowledgments:}  The authors would like to thank G Y Ma, S L Zhang, S Wang and Q Zhang  for helpful discussions. This research is supported by Natural Science Foundation of China with Project No. 11935007, 11805167.

\end{document}